\documentstyle[12pt]{article}
\newlength{\extraspace}
\newlength{\extraspaces}
\newcommand{\be}{\begin{equation}
\addtolength{\abovedisplayskip}{\extraspaces}
\addtolength{\belowdisplayskip}{\extraspaces}
\addtolength{\abovedisplayshortskip}{\extraspace}
\addtolength{\belowdisplayshortskip}{\extraspace}}
\newcommand{\ee}{\end{equation}}

\newcommand{\ba}{\begin{eqnarray}
\addtolength{\abovedisplayskip}{\extraspaces}
\addtolength{\belowdisplayskip}{\extraspaces}
\addtolength{\abovedisplayshortskip}{\extraspace}
\addtolength{\belowdisplayshortskip}{\extraspace}}
\newcommand{\ea}{\end{eqnarray}}

\newcommand{\newsection}[1]{
\vspace{15mm}
\pagebreak[3]
\addtocounter{section}{1}
\setcounter{equation}{0}
\setcounter{subsection}{0}
\setcounter{footnote}{0}
\begin{flushleft}
{\large\bf \thesection. #1}
\end{flushleft}
\nopagebreak
\medskip
\nopagebreak}

\newcommand{\Tr}{{\rm Tr}}
\newcommand{\Dmrns}{{\cal D}_{\mu\rho,\nu\sigma}}

\begin{document}

\addtolength{\baselineskip}{.8mm}

{\thispagestyle{empty}

\noindent \hspace{1cm}  \hfill HD--THEP--98--27 \hspace{1cm}\\
\mbox{}                 \hfill IFUP--TH 38/98 \hspace{1cm}\\
\mbox{}                 \hfill UCY--PHY--98/09 \hspace{1cm}\\
\mbox{}                 \hfill September 1998 \hspace{1cm}\\

\begin{center}
\vspace*{1.0cm}
{\large\bf Gauge--invariant quark--antiquark nonlocal condensates
in lattice QCD\footnote{Partially supported by MURST and by the EC TMR program
 ERBFMRX--CT97--0122.} }\\
\vspace*{1.0cm}
{\large M. D'Elia$^{\rm a}$, A. Di Giacomo$^{\rm b}$,
E. Meggiolaro$^{\rm c}$ }\\
%  \footnote{Postal address: Enrico Meggiolaro, Institut f\"ur Theoretische
%  Physik, Universit\"at Heidelberg, Philosophenweg 16, D--69120 Heidelberg,
%  Germany.
%  Fax nr.: +49--(0)6221--549333. 
%  E--mail address: E.Meggiolaro@thphys.uni-heidelberg.de .} 
\vspace*{0.5cm}{\normalsize
{$^{\rm a}$ Department of Natural Sciences,
University of Cyprus,
1678 Nicosia, Cyprus. \\
$^{\rm b}$ Dipartimento di Fisica,
Universit\`a di Pisa,
and INFN, Sezione di Pisa,
I--56100 Pisa, Italy. \\
$^{\rm c}$ Institut f\"ur Theoretische Physik,
Universit\"at Heidelberg,
Philosophenweg 16,
D--69120 Heidelberg, Germany.}}\\
\vspace*{2cm}{\large \bf Abstract}
\end{center}
\noindent
We study, by numerical simulations on a lattice, the behaviour of the
gauge--invariant quark--antiquark nonlocal condensates in the QCD vacuum
with dynamical fermions. A determination is also done in the {\it quenched}
approximation and the results are compared with the full--QCD case.
The fermionic correlation length is extracted and compared with the
analogous gluonic quantity.\\
\vspace{1.0cm}
\noindent
(PACS code: 12.38.Gc)
}
\vfill\eject

\newsection{Introduction}

\noindent
The so--called ``nonlocal condensates'', i.e., gauge--invariant field
correlators, are the original starting points for any standard calculation 
which adopts the method of QCD sum rules \cite{SVZ79}. 
These quantities appear when evaluating the power corrections, via the 
``Operator Product Expansion'' (OPE) \cite{Wilson69}, of the 
product of two hadronic currents. The effects due to the $x$--distribution
of these vacuum fluctuations sometimes have been neglected, 
dealing only with local condensates $\langle \bar{q}(0) q(0) \rangle$, 
$\langle G(0) G(0) \rangle$, etc.. However, it has been recognized 
in \cite{Gromes82,Mik-Rad86,Mik-Rad89,Rad91,Bak-Rad91,Mik-Rad92,Rad94}
that in many applications the effects due to 
the $x$--distribution of the nonlocal condensates have physical 
relevance and cannot be neglected.
Therefore, the knowledge of those nonlocal condensates from first principles
can be important for the study of the strong interaction 
theory and its applications. 

In a previous series of works, we have studied, by numerical simulations 
on a lattice, the gauge--invariant two--point correlators of the gauge
field strengths in the QCD vacuum:
\be
\Dmrns(x) = \langle 0| 
\Tr \left\{ G_{\mu\rho}(0) S(0,x) G_{\nu\sigma}(x) S^\dagger(0,x) \right\}
|0\rangle ~.
\ee
$G_{\mu\rho} = gT^aG^a_{\mu\rho}$ is the gauge field--strength tensor 
and $S(0,x)$ is the Schwinger phase operator needed to 
parallel--transport the tensor $G_{\nu\sigma}(x)$ to the point $0$.

These correlators have been determined on the lattice in the {\it quenched}
(i.e., pure gauge) theory, with gauge--group $SU(2)$ \cite{Campostrini84}, 
in the {\it quenched} $SU(3)$ theory in the range of physical distances between 
0.1 and 1 fm \cite{DiGiacomo92,npb97} and also in full QCD, i.e., including 
the effects of dynamical fermions \cite{plb97}.
The basic results of all these determinations is that the correlator
$\Dmrns(x)$, in the Euclidean theory, can be written as the sum of a 
perturbative--like term, behaving as $1/|x|^4$, and a nonperturbative
part, which falls down exponentially
\be
\Dmrns^{(n.p.)}(x) \sim \exp ( -|x|/\lambda_A ) ~.
\ee
The correlation length is $\lambda_A \simeq 0.13$ fm for the $SU(2)$ 
pure--gauge theory \cite{Campostrini84}, $\lambda_A \simeq 0.22$ fm for the 
$SU(3)$ pure--gauge theory \cite{DiGiacomo92,npb97} and $\lambda_A \simeq 0.34$
fm for full QCD (approaching the chiral limit) \cite{plb97}.

Along the same line, in this paper we present a lattice determination
of the quark--antiquark nonlocal condensates 
$\langle \bar{q}(0) S(0,x) q(x) \rangle$, adopting the same basic
strategies and techniques already developed for the study of the gluonic 
correlators \cite{DiGiacomo92,npb97,plb97}.
In particular, we shall make use of the ``cooling'' technique
\cite{Campostrini89,DiGiacomo90}
in order to remove the effects of short--range fluctuations on
large--distance correlators and get rid of the renormalizations.
We shall not present here again the details of our cooling procedure,
for which we refer the reader to our previous works 
\cite{DiGiacomo92,npb97,plb97,Campostrini89,DiGiacomo90}.

The quark--antiquark nonlocal condensates have been determined both in 
the $SU(3)$ pure gauge theory and in full QCD, i.e., including 
the effects of dynamical fermions. The details of the computations and the 
results are presented in Section 2. In Section 3 we conclude with some remarks 
about the results.

\newsection{Computations and results}

\noindent
As we shall explain in more detail later, for our lattice computations we
have used four flavours of {\it staggered} fermions, so that we are considering
a theory with four degenerate quark flavours in the continuum limit. 
Therefore, we have decided to
consider only the expectation values of those quark--antiquark operators
which are diagonal in flavour, but are nontrivial with respect of the
Dirac spinor indices, i.e.:
\be
C_i (x) = -\displaystyle\sum_{f=1}^4
\langle \Tr [\bar{q}^f_a (0) (\Gamma^i)_{ab} S(0,x) q^f_b (x)] \rangle ~.
\ee
A few words about the notation used in (2.1). $S(x,y)$ is the Schwinger 
string from $x$ to $y$
\be
S(x,y) \equiv {\rm P} \exp \left( i g \int_x^y dz^\mu A_\mu (z) \right) ~,
\ee
needed to make $C_i (x)$ a gauge--invariant object. ``P'' stands for
{\it ``path ordering''}: for simplicity, we take $S(x,y)$ along the 
straight--line path from $x$ to $y$. $A_\mu = A^a_\mu T^a$ is the
gauge field operator and $T^a$ are the matrices of the algebra of the 
colour group $SU(N_c)$ in the fundamental representation (For $N_c = 3$,
$T^a = \lambda_a/2$, where $\lambda_a$ are the Gell--Mann matrices).
The trace in (2.1) is taken with respect of the colour indices. 
The index $f$ in (2.1) is a flavour index ($f = 1,2,3,4$). 
The matrices $\Gamma^i$ are the sixteen independent $4 \times 4$ matrices of 
the Clifford's algebra acting on the Dirac indices $a,b$:
${\bf 1}$, $\gamma^\mu_E$, $\gamma^5_E$, $\gamma^\mu_E \gamma^5_E$,
$[\gamma^\mu_E,\gamma^\nu_E]/2$, where $\gamma^\mu_E$ ($\mu = 1,2,3,4$) 
and $\gamma^5_E$ are the Euclidean Dirac matrices. 

Making use of the invariance of the theory under P, T, translations and 
rotations, one can easily verify that most of the 
sixteen vacuum expectations values (2.1) are zero for symmetry reasons.
One is left with only two nontrivial quark--antiquark nonlocal condensates, 
out of the sixteen quantities $C_i (x)$ in (2.1), namely
\ba
C_0 (|x|) = -\displaystyle\sum_{f=1}^4 \langle \Tr [ \bar{q}^f_a (0)
S(0,x) q^f_a (x) ] \rangle ~; \nonumber \\
C_v (|x|) = {x_\mu \over |x|} C_\mu (x) =
-{x_\mu \over |x|}  \displaystyle\sum_{f=1}^4 \langle \Tr [ \bar{q}^f_a (0)
(\gamma^\mu_E)_{ab} S(0,x) q^f_b (x) ] \rangle ~.
\ea
For simplicity, in the following we shall refer to these two quantities as,
respectively, the {\it ``scalar nonlocal condensate''} [$C_0 (|x|)$] and
the {\it ``(longitudinal--) vector nonlocal condensate''} [$C_v (|x|)$].

In order to construct our operators on the lattice we have used the 
following procedure. In the {\it staggered} formulation 
quark fields live on elementary hypercubes, so our correlators can be defined
only for an even distance $d$ in lattice spacing units. In computing
$\langle \bar{q}(0) S(0,x) q(x) \rangle$ we have always put the point $0$ in 
the hypercube at the origin of the lattice, while $x$ has been varied
along the coordinate axes. 
The {\it staggered} propagators  $\langle \bar{\chi}_i \chi_j \rangle$  
have been computed at first and have been connected
to the origins of the hypercubes by minimal paths of gauge links
(an average over paths of equal length has been performed). They 
have then been combined to build up the quark propagator. Finally, the
Schwinger line connecting the origins of the two hypercubes has been put in.
We have also performed an average over different directions.

In this way we can construct two (adimensional) lattice operators
$C^L_0 (d \cdot a)$ and $C^L_v (d \cdot a)$ ($x = d \cdot a$, where $a$ is 
the lattice spacing and $d$ is the number of lattice spacings), which are 
proportional to $C_0 (d \cdot a)$ and $C_v (d \cdot a)$ respectively in the
na\"{\i}ve continuum limit, i.e., when the lattice spacing $a \to 0$:
\ba
C^L_0 (d \cdot a) \mathop\sim_{a\to0} a^3 
C_0 (d \cdot a) + {\cal O}(a^4) ~,\nonumber \\
C^L_v (d \cdot a) \mathop\sim_{a\to0} a^3 
C_v (d \cdot a) + {\cal O}(a^4) ~.
\ea
Higher orders in $a$ in (2.4) as well as possible multiplicative 
renormalizations are removed by cooling the quantum fluctuations at the scale 
of the lattice spacing, as explained in Refs. 
\cite{DiGiacomo92,npb97,plb97,Campostrini89,DiGiacomo90}. 
This removal will show up as a plateau in the dependence of the correlators 
on the number of steps of the cooling procedure: our data are the values of 
the correlators at the plateaux.

The computations have been performed with four flavours of {\it staggered}
fermions and the $SU(3)$ Wilson action for the pure--gauge sector: we have 
considered both the full--QCD case (i.e., including the effects of 
dynamical fermions) and the {\it quenched} case (where the effects coming 
from loops of dynamical fermions are neglected, i.e., ${\rm det} K[A] = 1$
in this approximation, $K[A]$ being the fermions matrix).

For the case with dynamical fermions we have measured the nonlocal 
condensates on a $16^3 \times 24$ lattice at $\beta = 5.35$ ($\beta =
6/g^2$, where $g$ is the coupling constant) and two different values of
the dynamical quark mass: $a \cdot m_q = 0.01$ and $a \cdot m_q = 0.02$.
In both cases the quark mass used in computing the quark propagator has been 
chosen to be the same one used during the simulation.
Our samples were $\sim 300$ configurations at $a \cdot m_q = 0.01$,
each one separated by 9 molecular dynamics time units,
and $\sim 70$ at $a \cdot m_q = 0.02$, each one separated by 12 molecular 
dynamics time units.
The computation was done on an QH4--APE machine by a standard 
hybrid Monte Carlo algorithm. 

For the {\it quenched} case the measurement has been performed on a $16^4$
lattice at $\beta = 6.00$, using a quark mass $a \cdot m_q = 0.01$ for 
constructing the external--field quark propagator, and also at $\beta = 5.91$, 
using a quark mass $a \cdot m_q = 0.02$. In both cases the value of $\beta$ 
was chosen in order to have the same physical scale as in full QCD
at the corresponding quark masses, thus allowing a direct comparison 
between the {\it quenched} and the full theory.
In this way we can see if the inclusion of dynamical fermions has or has not 
considerable effects on the quantities that we are measuring.

In the {\it quenched} case two other measurements have been done at 
$\beta = 6.00$, using quark masses $a \cdot m_q = 0.05$ and $0.10$ for 
constructing the external--field quark propagator, in order to study the 
dependence of the nonlocal condensates on the valence quark mass.

Finally, in all cases, we have measured the nonlocal condensates 
at distances $d = 2,4,6,8$ in units of lattice spacing.

In Figs. 1 and 2 we display the results for $a^3 C_0 (d \cdot a)$ and
$a^3 C_v (d \cdot a)$ respectively, versus the distance $d$ in lattice 
spacings, for full QCD at $a \cdot m_q = 0.01$ and
for full QCD at $a \cdot m_q = 0.02$.
The results for $a^3 C_0$ and $a^3 C_v$ obtained in the {\it quenched} 
determinations are displayed in Figs. 3 and 4, for the cases  $\beta = 6.00$ 
and $a \cdot m_q = 0.01$, $\beta = 5.91$ and $a \cdot m_q = 0.02$, 
$\beta = 6.00$ and $a \cdot m_q = 0.05$, $\beta = 6.00$ and 
$a \cdot m_q = 0.10$.

For the scalar nonlocal condensate we have tried a best fit to 
the data with the function
\be
C_0 (x) = A_0 \exp (-\mu_0 x) + {B_0 \over x^2} ~.
\ee
Results obtained in the various cases are shown in Table I.

Analogously, for the longitudinal--vector nonlocal condensate we have tried 
a best fit to the data with the function
\be
C_v (x) = A_v x^3 \exp (-\mu_v x) + {B_v \over x^3} ~.
\ee
Results obtained in the various cases are shown in Table II.

(The values of $\chi^2/N_{d.o.f.}$ reported in Tables I and II should be
considered as purely indicative of the goodness of our best fits,
since $N_{d.o.f.} \equiv n_{data} - n_{param.}$ is only an upper limit to 
the effective number of degrees of freedom after taking into account the 
correlations between data at different distances.)

The form of the perturbative--like terms in Eqs. (2.5) and (2.6) (i.e.,
$B_0/x^2$ for the scalar condensate and $B_v/x^3$ for the vector 
condensate) is that obtained in the leading order in perturbation theory, in 
the chiral limit $m_q \to 0$:
\ba
C^{(p.t.)}_0 (x) \mathop\sim_{m_q \to 0} \left( {m_q N_f N_c \over \pi^2}
\right) \cdot {1 \over x^2} ~; \nonumber \\
C^{(p.t.)}_v (x) \mathop\sim_{m_q \to 0} \left( {2 N_f N_c \over \pi^2}
\right) \cdot {1 \over x^3} ~,
\ea
for a theory with $N_f$ flavours and $N_c$ colours. In our case $N_f = 4$ 
and $N_c = 3$: the values for the coefficients $a B_0$ and $B_v$ obtained 
in the best fits agree as an order of magnitude with the estimates (2.7).

To conclude this section, let us make some remarks about the significance
of the results
obtained for the longitudinal vector nonlocal condensate. 
As the number of cooling steps tends to $\infty~$, the gauge--field
configuration is driven towards the zero--field configuration, 
so we expect both  
$C^L_0 (d)$ and $C^L_v (d)$ to reach asymptotically their 
free--field values. In both cases, and unlike the case of the gluon 
field--strength correlators, these free--field values are different from zero 
and correspond to the leading order in lattice perturbation theory.

We have determined the free--field values by measuring  
$C^L_0 (d)$ and $C^L_v (d)$ on the zero--field configuration.
Results are displayed in Figs. 5 and 6 respectively and compared
to results obtained in full QCD for $a \cdot m_q = 0.01$. 
Qualitatively similar figures are obtained for other values
of $a \cdot m_q$ and $\beta$.

It clearly appears that the vector nonlocal condensate
is strongly dominated by the free--field signal: our method to determine
it is thus likely to be poorly sensitive to the nonperturbative signal.
Moreover, the free--field signal for the vector nonlocal condensate is
only weakly mass--dependent (it is different from zero in the chiral
limit): this fact, together with the above--mentioned dominance of the
free--field signal, gives an explanation of why the data of the vector
nonlocal condensate, reported in Figs. 2 and 4, depend so weakly on the
value of the (valence) quark mass.
We regard results concerning the vector nonlocal condensate
as preliminary: it will be the subject of further investigations.

This is not the case of the scalar correlator: there the free--field 
signal is only a small fraction of the whole signal at most distances.
Moreover, as a further check of the reliability of the results
obtained in this case, we have tried to subtract the free--field 
signal from the measured scalar nonlocal condensate: we have obtained
that after the subtraction the results of the fits are unchanged for the 
nonperturbative part, while they give a value compatible with zero for the 
perturbative--like term.

\newsection{Discussion}

\noindent
In principle from our simulations we can extract two quantities of physical
interest: the quark condensate $\langle \bar{q}(0) q(0) \rangle$ and
the correlation length $\lambda_0 \equiv 1 / \mu_0$
of the scalar quark--antiquark nonlocal condensate. As explained in
Refs. \cite{Mik-Rad86,Mik-Rad89,Rad91,Bak-Rad91,Mik-Rad92,Rad94}, 
$\lambda_0$  plays a relevant role in many applications of QCD sum rules, 
especially for studying the pion form factors and the pion wave functions.

From the lattice we have extracted $\lambda_0$ in units of the lattice
spacing $a$. To convert from these units to physical units, the scale must
be set by comparison with some physical quantity. This is usually done by 
computing the string tension and/or the $\rho$ mass on the lattice and
comparing them with the physical values.
In the full--QCD case, on the $16^3 \times 24$ lattice at $\beta = 5.35$ 
with four flavours of {\it staggered} fermions, we have found,
by measuring $m_\pi$ and $m_\rho$ on our configurations and following 
the same procedure described in Ref. \cite{Laermann93}, the following values 
for the lattice spacing:
\ba
a (\beta = 5.35) &\simeq& 0.101~{\rm fm},~~{\rm for}~ a \cdot m_q = 0.01 ~;
\nonumber \\
a (\beta = 5.35) &\simeq& 0.120~{\rm fm},~~{\rm for}~ a \cdot m_q = 0.02 ~.
\ea
In the {\it quenched} case the lattice spacing is approximately \cite{Boyd96}:
\ba
a^{(YM)} (\beta = 6.00) &\simeq& 0.103~{\rm fm}~;
\nonumber \\
a^{(YM)} (\beta = 5.91) &\simeq& 0.120~{\rm fm}~.
\ea
Using these values, we can extract the physical values
of the correlation length $\lambda_0$ for all the cases examined, obtaining
the results reported in Table III.

At $a \cdot m_q = 0.01$ the value is roughly twice as big as the value for 
the correlation length $\lambda_A$ of the gluon field strength, defined in 
Eq. (1.2), at the same quark mass $a \cdot m_q = 0.01$ \cite{plb97}.

The full--QCD and the {\it quenched} correlation lengths $\lambda_0$ are 
nearly the same, when compared at the same quark mass. 

Similarly to the gluon case \cite{plb97}, the fermionic correlation length 
appears to decrease when increasing the quark mass. However, our results seem 
to suggest that the value of $\lambda_0$ is sensitive to the value of the 
valence quark mass used in constructing the quark propagator, but is not
much influenced by the dynamical quark loops coming from the determinant of
the fermionic matrix ${\rm det} ( K[A] )$.

Using the values of the pion mass $m_\pi$, measured on our configurations
in the full--QCD case [$a \cdot m_\pi = 0.263(8)$ for $a \cdot m_q = 0.01$,
and $a \cdot m_\pi = 0.372(13)$ for $a \cdot m_q = 0.02$], and the
corresponding values of $a \cdot \mu_0$ reported in Table I, we
find that $m_\pi/\mu_0 = 1.6(4)$ for $a \cdot m_q = 0.01$ and
$m_\pi/\mu_0 = 1.4(2)$ for $a \cdot m_q = 0.02$. Therefore, the inverse of 
the scalar correlation length, $\mu_0 = 1/\lambda_0$, turns out to be 
proportional (within the errors) to the pion mass $m_\pi$: in other
words, it is the pion mass $m_\pi$ which determines the length--scale
of the scalar nonlocal condensate.

We conclude with a brief comment about the quark condensate. A way to extract
this quantity is to consider the uncooled values of the scalar quark 
correlator at zero distance, $C^L_0(0)$, for different quark masses and
then to extrapolate those values to zero quark mass. From our full--QCD
simulations we have obtained the following results (in lattice units):
\ba
C^L_0(0) &=& 7.17(6) \times 10^{-2},~~{\rm for}~ a \cdot m_q = 0.01 ~;
\nonumber \\
C^L_0(0) &=& 1.26(2) \times 10^{-1},~~{\rm for}~ a \cdot m_q = 0.02 ~.
\ea
A linear extrapolation of these two values to zero quark mass leads to the
result: 
\be
C^L_0(0)|_{m_q = 0} = 0.0174(32) ~.
\ee
Adopting the same procedure outlined in Ref. \cite{Laermann93} (which
properly takes into account the anomalous dimension of the quark condensate),
we can then extract the following value for the single--flavour quark
condensate in the $\overline{\rm MS}$ renormalization scheme at a scale of
$\mu = 1$ GeV:
\be
|\langle \bar{u} u \rangle|^{(\overline{\rm MS})} (\mu = 1 ~{\rm GeV}) =
0.013(2) ~{\rm GeV}^3 ~.
\ee
This results is in perfect agreement with the phenomenological value
\cite{Dosch96}.

As an alternative method, one could also try to extract the quark condensate
directly from the values of the coefficient $A_0$, defined in Eq. (2.5).
However, the values of $a^3 A_0$ reported in Table I in the full--QCD case
for $a \cdot m_q = 0.01$ and $a \cdot m_q = 0.02$ are visibly much smaller
than the corresponding values for $C^L_0(0)$ reported in Eq. (3.3).
Apparently, no reasonable quark condensate can be extracted from these
values of $a^3 A_0$.
The reason for this discrepancy could lie in the anomalous dimension of
the quark condensate: while we believe that the correlation length 
$\lambda_0$ is not affected by the cooling procedure, we do not know the
effects of cooling on the quark condensate, which has an anomalous
dimension. (This problem is not present in the case of gluon correlators,
since the extracted gluon condensate is renormalization group invariant.)
We hope to come back to this point in future works.

\vfill\eject

{\renewcommand{\Large}{\normalsize}
}

\vfill\eject

\noindent
\begin{center}
{\bf TABLE CAPTIONS}
\end{center}
\vskip 0.5 cm
\begin{itemize}
\item [\bf Tab.~I.] Results obtained from a best fit to the data of the
scalar nonlocal condensate with the function (2.5), in the various cases
that we have examined (``$f$'' stands for {\it ``full--QCD''}, while
``$q$'' stands for {\it ``quenched''}).
\bigskip
\item [\bf Tab.~II.] Results obtained from a best fit to the data of the
longitudinal--vector nonlocal condensate with the function (2.6), in the 
various cases that we have examined (``$f$'' stands for {\it ``full--QCD''}, 
while ``$q$'' stands for {\it ``quenched''}).
\bigskip
\item [\bf Tab.~III.] The physical values of the correlation length 
$\lambda_0$ for all the cases that we have examined. Reported errors
refer only to our determination and do not include the uncertainty on 
the physical scale.
\end{itemize}

\vfill\eject

\newpage

\vskip 2cm

\centerline{\bf Table I}

\vskip 5mm

\moveright 0.25 in
\vbox{\offinterlineskip
\halign{\strut
\vrule \hfil\quad $#$ \hfil \quad & 
\vrule \hfil\quad $#$ \hfil \quad & 
\vrule \hfil\quad $#$ \hfil \quad & 
\vrule \hfil\quad $#$ \hfil \quad & 
\vrule \hfil\quad $#$ \hfil \quad & 
\vrule \hfil\quad $#$ \hfil \quad \vrule \cr
\noalign{\hrule}
\beta, ~{\rm theory} &
a \cdot m_q &
a^3 A_0 \times 10^2     &
a \mu_0     &
a B_0  \times 10^1 &
\chi^2/N_{d.o.f.}   \cr
& & & & & \cr
\noalign{\hrule}
\noalign{\hrule}
5.35, ~{\rm f}  & 0.01  & 0.49(13) & 0.16(4) & 0.13(3) & 1.3\cdot 10^{-2}\cr
\noalign{\hrule}
5.35, ~{\rm f} & 0.02 & 1.7(5) & 0.26(4) & 0.19(10) & 6.4\cdot 10^{-3} \cr
\noalign{\hrule}
6.00, ~{\rm q} & 0.01 & 1.6(5) &0.16(4) & 0.09(12) & 7.6 \cdot 10^{-2} \cr
\noalign{\hrule}
5.91, ~{\rm q}  & 0.02 & 2.3(7)   & 0.26(3)  & 0.25(14) & 5.2 \cdot 10^{-2} \cr
\noalign{\hrule}
6.00, ~{\rm q} & 0.05 & 1.8(4)  & 0.34(2) & 0.7(1) & 0.2 \cr
\noalign{\hrule}
6.00, ~{\rm q} & 0.10 & 5.6(5) & 0.55(1) & 1.0(1) & 1.3 \cdot 10^{-2} \cr
\noalign{\hrule}
}}

\vskip 2cm

\centerline{\bf Table II}

\vskip 5mm

\moveright 0.37 in
\vbox{\offinterlineskip
\halign{\strut
\vrule \hfil\quad $#$ \hfil \quad & 
\vrule \hfil\quad $#$ \hfil \quad & 
\vrule \hfil\quad $#$ \hfil \quad & 
\vrule \hfil\quad $#$ \hfil \quad & 
\vrule \hfil\quad $#$ \hfil \quad & 
\vrule \hfil\quad $#$ \hfil \quad \vrule \cr
\noalign{\hrule}
\beta, ~{\rm theory} &
a \cdot m_q &
a^6 A_v  &
a \mu_v     &
B_v  &
\chi^2/N_{d.o.f.}   \cr
& & & & & \cr
\noalign{\hrule}
\noalign{\hrule}
5.35, ~{\rm f}  & 0.01  & 0.19(1) & 1.46(1) & 2.87(3) & 1.76 \cr
\noalign{\hrule}
5.35, ~{\rm f} & 0.02 & 0.20(1) & 1.48(2) & 2.84(3) & 0.8 \cr
\noalign{\hrule}
6.00, ~{\rm q} & 0.01 & 0.21(1) & 1.48(1) & 2.83(2) & 0.4 \cr
\noalign{\hrule}
5.91, ~{\rm q}  & 0.02 & 0.216(7) & 1.494(6) & 2.81(1) & 2.3 \cdot 10^{-3} \cr
\noalign{\hrule}
6.00, ~{\rm q} & 0.05 & 0.22(1)  & 1.50(1) & 2.80(1) & 2 \cdot 10^{-2} \cr
\noalign{\hrule}
6.00, ~{\rm q} & 0.10 & 0.26(1) & 1.54(1) & 2.73(2) & 2.2 \cr
\noalign{\hrule}
}}

\vskip 2cm

\centerline{\bf Table III}

\vskip 5mm

\moveright 1.8 in
\vbox{\offinterlineskip
\halign{\strut
\vrule \hfil\quad $#$ \hfil \quad & 
\vrule \hfil\quad $#$ \hfil \quad & 
\vrule \hfil\quad $#$ \hfil \quad \vrule \cr
\noalign{\hrule}
\beta, ~{\rm theory} &
a \cdot m_q &
\lambda_0 ~({\rm fm})   \cr
& & \cr
\noalign{\hrule}
\noalign{\hrule}
5.35, ~{\rm f}  & 0.01  & 0.63^{+0.21}_{-0.13} \cr
\noalign{\hrule}
5.35, ~{\rm f} & 0.02 &  0.46^{+0.09}_{-0.06} \cr
\noalign{\hrule}
6.00, ~{\rm q} & 0.01 & 0.64^{+0.22}_{-0.13} \cr
\noalign{\hrule}
5.91, ~{\rm q}  & 0.02 & 0.46^{+0.06}_{-0.05} \cr
\noalign{\hrule}
6.00, ~{\rm q} & 0.05 & 0.30(2) \cr
\noalign{\hrule}
6.00, ~{\rm q} & 0.10 & 0.187(3) \cr
\noalign{\hrule}
}}

\vfill\eject

\noindent
\begin{center}
{\bf FIGURE CAPTIONS}
\end{center}
\vskip 0.5 cm
\begin{itemize}
\item [\bf Fig.~1.] The function $a^3 C_0(x)$ (scalar correlator)
versus the distance $d = x/a$ in lattice spacings, for the full--QCD case at
$\beta = 5.35$ and quark masses $a \cdot m_q = 0.01$ (circles) and
$a \cdot m_q = 0.02$ (squares). 
The curves correspond to our best fits [Eq. (2.5)].
\bigskip
\item [\bf Fig.~2.] The function $a^3 C_v(x)$ (vector correlator)
versus the distance $d = x/a$ in lattice spacings, for the full--QCD case at
$\beta = 5.35$ and quark masses $a \cdot m_q = 0.01$ (circles) and
$a \cdot m_q = 0.02$ (squares). This last set of symbols (squares) has been 
shifted horizontally in the right direction to distinguish it from the other 
set of symbols (circles). As an example, we plot the curve corresponding to 
our best fit [Eq. (2.6)] to the data at $a \cdot m_q = 0.01$.
\bigskip
\item [\bf Fig.~3.] The same as in Fig. 1 for the {\it quenched} case
at $\beta = 6.00$ and a quark mass $a \cdot m_q = 0.01$ (circles),
$\beta = 5.91$ and $a \cdot m_q = 0.02$ (squares),
$\beta = 6.00$ and $a \cdot m_q = 0.05$ (triangles down),
$\beta = 6.00$ and $a \cdot m_q = 0.10$ (triangles up).
\bigskip
\item [\bf Fig.~4.] The same as in Fig. 2 for the {\it quenched} case
at $\beta = 6.00$ and a quark mass $a \cdot m_q = 0.01$ (circles),
$\beta = 5.91$ and $a \cdot m_q = 0.02$ (squares),
$\beta = 6.00$ and $a \cdot m_q = 0.05$ (triangles down),
$\beta = 6.00$ and $a \cdot m_q = 0.10$ (triangles up).
Again, different sets of symbols have been shifted horizontally in the
right direction. As an example, we plot the curve corresponding to 
our best fit [Eq. (2.6)] to the first set of data (circles).
\bigskip
\item [\bf Fig.~5.] The free--field value of $C^L_0 (d)$, measured on the 
zero--field configuration (filled circles), compared to the results obtained
in full QCD at $a \cdot m_q = 0.01$ (open circles).
\bigskip
\item [\bf Fig.~6.] The free--field value of $C^L_v (d)$, measured on the 
zero--field configuration (filled circles), compared to the results obtained
in full QCD at $a \cdot m_q = 0.01$ (open circles).
\end{itemize}

\vfill\eject

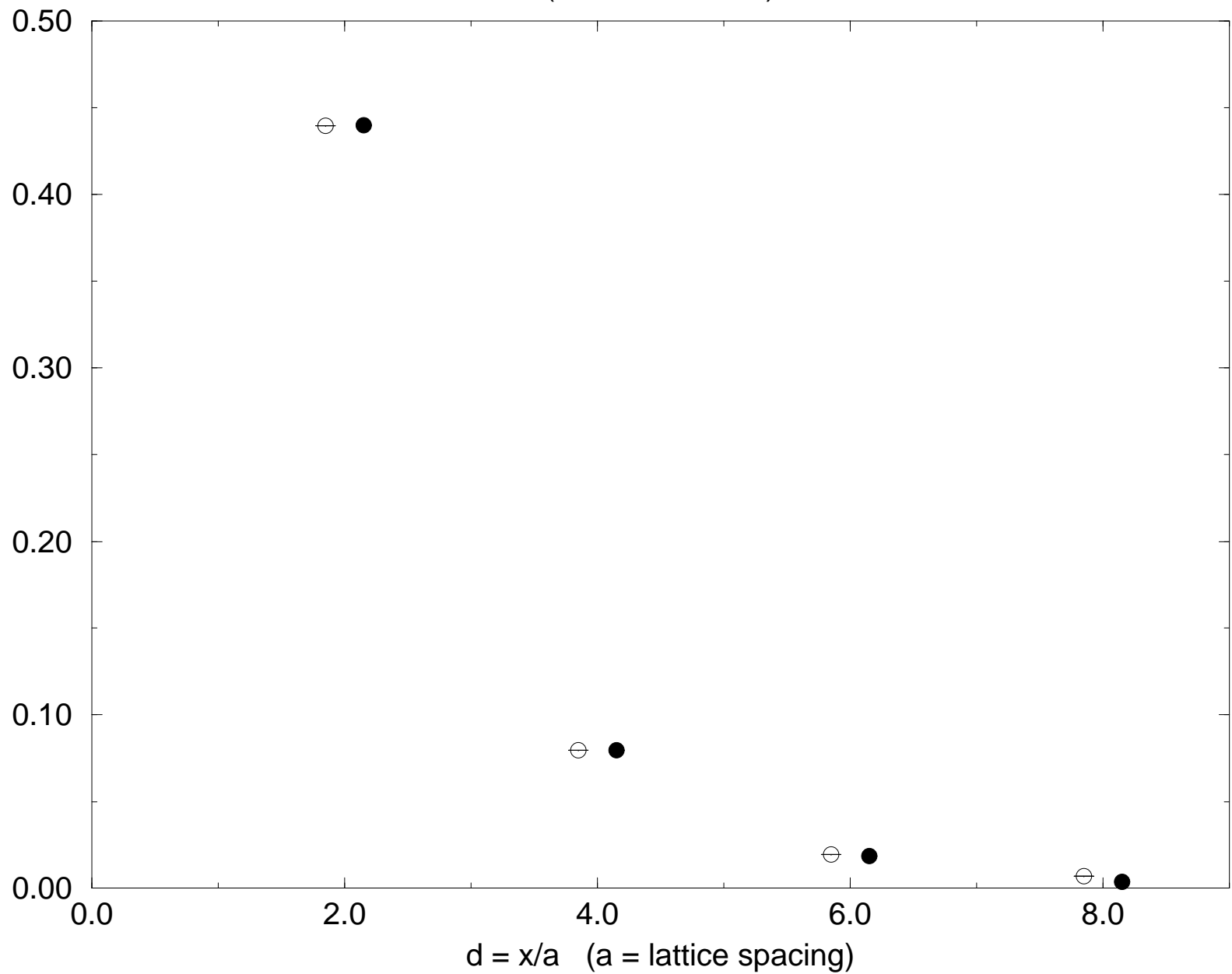
\end{document}